\documentclass[letterpaper, 10pt, journal,comsoc]{IEEEtran}

\usepackage{blindtext, graphicx, amsmath, amsfonts, amssymb}

\usepackage{microtype} % Fine-tunes the spacing between letters, words and lines
\usepackage{cite}
\usepackage{url}
\usepackage{hyperref}
\usepackage[table,svgnames]{xcolor} 
\newcommand{\red}[1]{{\color{black}#1}}
\usepackage{epstopdf}

\usepackage{multirow}

\begin{document}

%\relscale{1}
%\sloppy                % Some small and clever fine-tuning for the line breaks
%\relscale{0.97}        % Changes the font size to 99% of the original font size

\title{Positioning and Location-Aware Communications for Modern Railways with 5G New Radio}
% \title{Iterative Refinement Algorithms for Joint Position and Orientation Estimation in 5G mm-Wave}

\author{
\IEEEauthorblockN{Jukka Talvitie, Toni Levanen, Mike Koivisto, Tero Ihalainen, Kari Pajukoski, and Mikko Valkama %
}
\thanks{J. Talvitie, T. Levanen, M. Koivisto, and M. Valkama are with Tampere University, Finland (firstname.lastname@tuni.fi)}
\thanks{T. Ihalainen and K. Pajukoski are with Nokia Bell Labs, Finland (firstname.lastname@nokia-bell-labs.com)}
\thanks{This article contains multimedia material, available at \newline {\color{blue}\texttt{\href{http://www.tut.fi/5G/HST_positioning/}{http://www.tut.fi/5G/HST\_positioning/}}}}
\thanks{\copyright~2019 IEEE.  Personal use of this material is permitted.  Permission from IEEE must be obtained for all other uses, in any current or future media, including reprinting/republishing this material for advertising or promotional purposes, creating new collective works, for resale or redistribution to servers or lists, or reuse of any copyrighted component of this work in other works.}
}

\maketitle

%%%%%%%%%%%%%%%%%%%%% Abstract %%%%%%%%%%%%%%%%%%%%%%
\begin{abstract}
Providing high-capacity radio connectivity for high-speed trains (HSTs) is one of the most important use cases of emerging 5G New Radio (NR) networks. In this article, we show that 5G NR technology can also facilitate high-accuracy continuous localization and tracking of HSTs. Furthermore, we describe and demonstrate how the NR network can utilize the continuous location information for efficient beam-management and beamforming, as well as for downlink Doppler precompensation in the single-frequency network context. Additionally, with particular focus on millimeter wave networks, novel concepts for low-latency intercarrier interference (ICI) estimation and compensation, due to residual Doppler and oscillator phase noise, are described and demonstrated. The provided numerical results at 30 GHz operating band show that sub-meter positioning and sub-degree beam-direction accuracies can be obtained with very high probabilities in the order of 95-99\%. The results also show that the described Doppler precompensation and ICI estimation and cancellation methods substantially improve the throughput of the single-frequency HST network. 

%\red{TEXT HERE}

\end{abstract}


\section{Introduction}
\label{sec:introduction}

\begin{figure*}[t]
    \centering
    \includegraphics[width=0.8\textwidth]{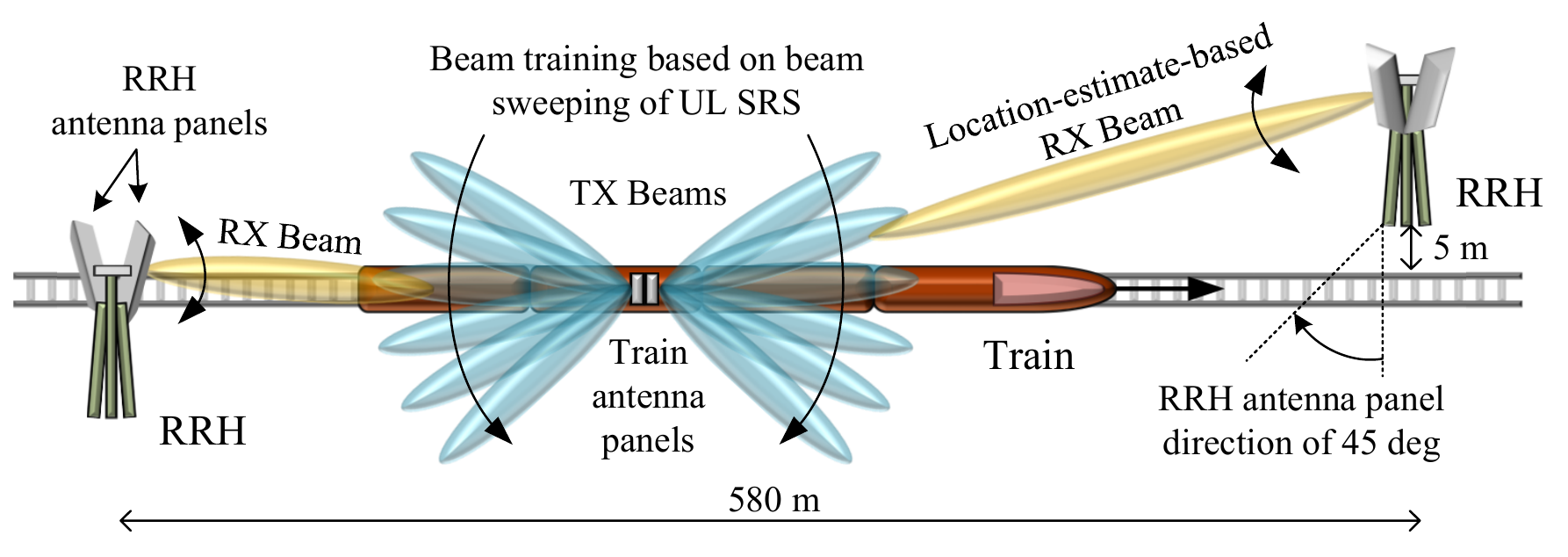} %\includegraphics[width=0.25\textwidth, angle=-90]{system_illust2.eps}
    \caption{Illustration of the considered railway scenario and the single-frequency mmWave radio network composing of multiple remote radio heads (RRHs).}
    \label{fig:system}
\end{figure*}

%\red{
% \begin{itemize}
%     \item Background and motivation 
%     \item 5G-enabled opportunities for modern railways (focus on aspects given in the Magazine call)
%     \item State-of-the-art (pos + comms)
%     \item Considering the 3GPP-specified HST-scenario as a baseline (i.e. emphasizing that the studied scenario is generally recognized by the community)
%     \item A few words about the status of standardization \cite{HST_standardization_magazine_article}.
%     \item Aim of the article
% \end{itemize}
% }

The upcoming 5G new radio (NR) networks have the potential to revolutionize the service and management opportunities of modern railway systems by introducing flexible and high-performance communications and positioning capabilities \cite{HST_standardization_magazine_article,development_trends_railway_comms}. Compared to previous mobile network generations, high-accuracy radio positioning is generally considered as one of the most exciting new key features of 5G NR networks, and it is expected that submeter or even centimeter-level positioning accuracy with high reliability can be provided, especially in the networks operating at the mmWave bands \cite{Henk_vehicular_pos, mike_5G_pos}. Moreover, besides providing extreme broadband connectivity for increased data rates, 5G NR networks support mission-critical use cases with predefined requirements for the communications link performance, such as end-to-end latency, availability and reliability \cite{5G_overview}.

Whether the objective is to offer extremely high data rate mobile connections for passengers, or to support ultra-reliable and low-latency mission-critical railway management functionalities, 5G NR networks are able to provide diverse set of services via network slicing \cite{HST_standardization_magazine_article}. For each desired railway system functionality, a unique network slice can be dedicated to support a specific set of required key performance indicators. Moreover, as 5G networks are able to offer multi-purpose services based on a single physical network, the implementation and maintenance cost of railway systems can be reduced compared to compound solutions where separate operations are distributed among multiple dedicated systems.

Due to the above-described great potential of 5G NR networks in modern railway systems, high-speed train (HST) aspects have been clearly recognized also in the 3rd generation partnership project (3GPP) standardization, see, e.g., \cite{TR_38913_scenario,HST_standardization_magazine_article}. To this end, a baseline 5G NR network deployment and system parameterization for HST radio access research and performance evaluations, as illustrated at a conceptual level in Fig. \ref{fig:system}, has been defined in \cite{TR_38913_scenario}. Nonetheless, detailed requirements for railway communications, discussed, e.g., in \cite{Sand_paper_railways}, are still under the specification stage, but it is clear that both mission-critical use cases, as well as extreme broadband use cases with high data rates, are considered.

%\footnotemark

%\footnotetext{See http://www.3gpp.org/news-events/partners-news/1964-frmcs\_r16}
% and 3GPP TS 22.289

In this article, we describe a novel 5G positioning and communications concept for modern railway systems based on the 5G NR specification guidelines operating at the mmWave bands. The developed high-accuracy radio positioning scheme introduces considerable benefits for modern railway systems, especially including railway management functionalities and other mission critical services. \red{As a standalone technology, 5G NR based positioning and tracking increase the diversity of the overall multitechnology train positioning solution, and thus improves the reliability and availability of the position information compared to current positioning solutions. It is generally known that, e.g., global navigation satellite systems (GNSS) alone are not able to provide the required positioning performance, and thus fusion over multiple positioning technologies, including the proposed standalone 5G positioning approach, is necessary to fulfill the required performance criteria \cite{multisensor_train_pos,ETCS_odometry_req}}. Besides describing and demonstrating the 5G NR based train positioning and tracking, we also discuss the utilization of the continuous position information in the NR data channel. Specifically, we demonstrate the feasibility of the location-based beamforming at the network side to enhance the efficiency of beam-alignment and beam-tracking in the mmWave HST network via reduced overhead and latency compared to conventional beam training procedures. Furthermore, with HSTs, the signal degradation due to high Doppler shifts at the mmWave bands can significantly limit the system performance, particularly in the single-frequency network (SFN) context where signals from multiple transmission points or remote radio heads (RRH) are superimposed at the receiver. However, presuming access to real-time position and velocity information, we further demonstrate that appropriate Doppler pre-compensation methods can be introduced to alleviate the signal degradation due to high mobility. Finally, an important implementation challenge in the mmWave frequencies, namely the oscillator phase noise (PN), is considered by introducing novel signal processing solutions for mitigating the oscillator PN effects in the mmWave 5G NR networks.

\section{5G NR Prospects for Modern Railways: Positioning and Communications}
\label{sec:5GProspects}

\subsection{System Description}

Building on the 3GPP vision of 5G communications for HSTs presented in \cite{TR_38913_scenario}, the railway scenario considered in this article is illustrated in Fig. \ref{fig:system}. The train is equipped with a relay node including two antenna panels directed towards the nose and tail of the train. All communications between the network and the train is routed via the train relay, such as passenger communications and dedicated communication links of the underlying railway management system. Thus, instead of managing frequent and fast handovers for hundreds of passengers simultaneously, only a single relay node is utilized \cite{TR_38913_scenario}. This reduces the network interference and increases the achieved data throughput due to reduced control signaling overhead. By following the 3GPP guidelines, the network consists of RRHs, whose locations are alternating between both sides of the track with inter-site distance of 580~m\red{, and with 5~m distance to the track}. Each RRH is equipped with two antenna panels, rotated by 45 degrees with respect to the track normal, covering both track directions.

The network is assumed to operate at 30 GHz mmWave band according to the SFN principle, where multiple RRHs transmit to or receive from the train relay using the same time-frequency resources. %Therefore, depending on the network configurations, a predefined set of RRHs transmit the same downlink signal, which is received by the train panels. %Correspondingly, the uplink signals transmitted by the train are received at multiple RRHs at the same time. %Due to the included radio wave propagation delays, a sufficient timing alignment for the transmissions between the train and the RRHs should be considered in order to avoid inter-symbol interference between consecutive OFDM symbols. However, 
In downlink, network side time alignment is used to align the received signals in the train within the used \red{cyclic prefix (CP)} length. Since the RRH signals are mostly received from the angles corresponding approximately to the direction of the nose and tail of the train, separate antenna panels at the train observe significantly different Doppler shifts. This causes considerable degradation of the radio link quality, and therefore it is important to introduce appropriate Doppler shift precompensation, as discussed and demonstrated later in this article. %Interestingly, because the observed Doppler shift depends on the velocity and the underlying system geometry, the performance of the precompensation method is highly dependent on the accuracy of the proposed train positioning approach.

\subsection{5G Positioning Prospects for Modern Railways}
\label{sec:5GProspectsPos}

Positioning is, in general, considered as one of the key features in the upcoming 5G networks and various positioning use cases with related performance requirements have been specified in detail by the 3GPP \cite{TR_22872_pos_requrements}. Besides focusing only on pursuing high positioning accuracy, also other performance indicators are considered including availability, latency, reliability, and integrity of the position estimates. In addition to actual positioning, estimation of velocity and device bearing are also addressed. Moreover, by exploiting the flexibility, versatility and configurability of the 5G-enabled positioning services, the positioning system can be customized for specific needs of modern railway systems. Thus, together with high-performance communications capability, 5G networks introduce a great asset for railway system management in terms of efficiency, affordability and safety.

Besides being an essential part of railway system management, positioning can benefit the performance of the 5G communications itself \cite{C:2017:Levanen:LocationAware5GComms}. In location-aware communications, position information can be utilized for a vast set of location-aware \red{radio resource management (RRM)} functionalities, such as location-based beamforming and proactive resource allocation. In addition, position information can be used for reducing the overhead of beam training processes, and for mitigating the signal impairment due to Doppler phenomenon in high %velocity use cases, such as with high-speed trains.
mobility use cases.

In general, compared to previous mobile network generations, 5G networks benefit from the large signal bandwidths available at the mmWave bands, which enable very accurate ranging measurements for positioning systems \cite{Henk_vehicular_pos}. In order to compensate for the increased path losses at the mmWave bands, large antenna arrays with effective beamforming capability are also introduced. From the positioning perspective, this means a potential access to high-accuracy angle measurements, and thus for improved positioning performance. Furthermore, line-of-sight (LOS) links are mostly available, which offers a straightforward utilization of both the ranging-based measurements and the angle-based measurements in various positioning algorithms, particularly in outdoor scenarios \cite{Henk_vehicular_pos, mike_5G_pos}.

Although 5G networks have evidently potential to reach submeter positioning accuracies, each positioning scenario has to be separately considered based on the given performance requirements \cite{TR_22872_pos_requrements}. In the studied railway scenario, there are several challenges, which have to be addressed in order to meet the given performance requirements. Firstly, the expected high velocities necessitate low latency and high positioning update rate in order to support mission critical railway management. \red{Because of the expectedly large mass of a train, high-accuracy predictions of the train position and corresponding train dynamics are often feasible over a limited time period. However, with considerable track curvature and varying train acceleration, the prediction error can accumulate rapidly over given tolerable accuracy levels.} Secondly, the specific access node geometry of the considered railway scenario is rather challenging, as the RRHs are aligned approximately on the same line with the train. This causes degraded geometric dilution of precision, which limits the achievable positioning performance. Finally, when considering ranging measurements attained by taking timing measurements of the radio wave propagation delays, the clock offsets between the train and the network nodes have to be taken into account. In the position approach described in Section \ref{sec:5GPos}, these aspects are properly reflected.

%\red{ToLe: Should we comment clock offset between network elements here also?}

\subsection{5G Communications Prospects for Modern Railways}
\label{sec:5GProspectsComms}

Similar to the positioning, 5G communications at the mmWave bands benefits from the large available bandwidth, which increases the system capacity as well as provides means for enhanced transmission diversity and supplementary degrees of freedom in data scheduling. The role of beamforming has significantly increased in 5G NR, compared to Long Term Evolution based systems, and even the regularly broadcasted synchronization signals and fundamental system information are beamformed in the form of synchronization signal blocks \cite{TS_38211_signal}. Furthermore, diverse beamforming techniques provide effective utilization of spatial domain via large antenna arrays, and are considered as one the most promising techniques for reducing network interference, which is especially crucial in heterogeneous network deployments with highly dynamic time-division duplex %(TDD) 
transmissions. Another important feature of 5G NR is the flexibility and configurability of the air interface, particularly in the form of the scalable subcarrier spacing, which significantly increases the tolerance against substantial Doppler spreads in high mobility scenarios. %Currently, the 5G NR supports subcarrier spacings of 15~kHz, 30~kHz, and 60~kHz for below 6~GHz carrier frequencies, and 60~kHz and 120~kHz for mmWave bands. Also, 240~kHz subcarrier spacing is supported for synchronization signal blocks to reduce the time required by beam training. 

% By following the guidelines of the 3GPP railway scenario \cite{TR_38913_scenario}, the network operates according to a single frequency principle, where transmissions from separate RRHs are performed simultaneously by using the same time-frequency resources. Moreover, assuming the proposed two antenna panel design for the train, it should be noted that the signals received from the opposite directions have opposite signed Doppler shifts. This leads to severe problems 

The SFN operating principle of the HST network provides good macro diversity, but has also certain technical challenges. Particularly, due to largely different Doppler shifts of the signals coming from different RRHs, there can be severe problems in channel estimation and received signal quality, especially if it is assumed that the received signals from the different panels are combined to be processed with a single baseband to reduce the cost of the implementation. Therefore, a network side Doppler precompensation technique is proposed in this article. The idea is to estimate the Doppler shift per communications link based on the estimated train position and velocity and known RRH positions. With this information, the network can calculate estimates of link-wise Doppler shifts and thereon pre-compensate the transmit signals per RRH. This leads to more stable LOS channel experienced through separate communication links and allows to combine signals in the radio frequency system before analog-to-digital conversion and baseband processing in the train receiver. This technique and its benefits are evaluated and demonstrated in Section \ref{sec:5GComms}.

% Note: references to other sections are possibly not considered in the final version (should be something like this:"are evaluated later on in this article"

Another potentially significant error source in a SFN is the timing alignment of the downlink signals. By timing alignment we refer to the relative timing of the received signals. When the train is receiving signals from two or more RRHs, time difference of the arriving waveforms is defined by the difference in distance to the serving RRHs. %For example, when the train is exactly in the middle of two RRHs, the signals are received with perfect alignment. However, when the train is close to one of the RRHs, the distances to the other RRHs are significantly larger and can cause significant interference in the receiver. %With an example cyclic prefix length of 586~ns \cite{TR_38913_scenario}, the maximum difference in the distance of the serving RRHs, in order to align the received signals to arrive within the cyclic prefix, is only 176~m.
For this reason, so called network timing alignment is also proposed and considered for high speed single frequency networks%. The network timing alignment uses 
, where the estimated train position and known RRH positions are used to calculate the time it takes for the transmitted signals to reach the receiver. %Then, on the RRH transmitters, the transmitted signals are delayed to allow for alignment of all transmitted signals in the train receiver. %This technique and its benefits are evaluated in the Section \ref{sec:5GComms}. 

%In addition, the relatively large subcarrier spacings supported by 5G NR in the mmWave band alleviate the effects of PN. At higher carrier frequencies
In mmWave communications, PN is typically more pronounced limiting the use of smaller subcarrier spacings and/or higher-order modulation and coding schemes (MCSs). With CP orthogonal frequency division multiplexing (CP-OFDM), PN causes common phase error (CPE) and inter-carrier interference (ICI). The CPE is observed as a common phase rotation over all active subcarriers and can be relatively easily estimated and compensated. For this reason, in 5G NR Rel-15, a new reference signal type called phase tracking reference signal (PTRS) was introduced to allow estimation and compensation of CPE in mmWave communications. The ICI, in turn, typically impairs the performance with higher MCSs. %, especially at high velocities due to increased Doppler shift or at mmWave communications due to increased PN distortion. 
The current PTRS design of 5G NR Rel-15 does not support estimating or compensating the ICI. In Section \ref{sec:5GComms}, we will demonstrate a novel block based PTRS design \cite{C:2017:Levanen:LocationAware5GComms}, \cite{VillesPhaseNoisePaper} and show its benefits on the link throughput in the considered %high speed train 
mmWave HST scenario.

%\red{Regarding the high mobility of the trains and the typical system geometry illustrated in Fig. \ref{fig:system}, the beam management becomes non-trivial in when the train passes RRHs. The narrower the used beams are and the closer the RRHs are located to the train track, the faster is the beam switching rate when the train passes a serving RRH. This may lead to interruptions in the high data rate communications link. Therefore, sophisticated solutions to note the high beam switching rate near by RRHs are required to diminish the throughput losses when passing a serving RRH and switching the serving RRHs in the different train panels. The two panel design assumed for the train%, where two antenna panels are pointing either towards the direction of travel or to the opposite direction, 
%allows the train to maintain simultaneous connection to two RRHs. This allows to maintain connectivity when passing RRHs and also improves the received signal power when the train is traveling between two RRHs.}

% \red{ToLe: IF I'm able to create example results of the TA and Doppler precompensation effect, we could add the example figure here to support this discussion and reserve the Section \ref{sec:5GComms} for throughput results with all the bells and whistles..}

\section{5G NR based Positioning of High-Speed Trains in Modern Railways}
\label{sec:5GPos}

To benefit from the large bandwidths of 5G NR, and to avoid tight requirements on the clock synchronization between the train and the network, the considered positioning approach relies on uplink time-difference-of-arrival (TDOA) measurements at the RRHs. Since the measurements and the related positioning algorithms are managed at the network side, the network possesses always the most recent position information, which is crucial for the operation of location-aware RRM and low-latency railway management systems. The used TDOA measurements are based on the uplink sounding reference signal (SRS), specified for the 5G NR in \cite{TS_38211_signal}. According to the TDOA-based positioning principle, the RRHs are assumed mutually time-synchronized, which is well justified especially when the RRHs are under the same baseband unit. However, clock synchronization between the network and the train is not required, as the TDOA processing effectively removes any clock offsets between the transmit and receive nodes.

%\vspace*{-1\baselineskip}

\subsection{TDOA/EKF based Positioning Engine}
\label{sec:TDOA_EKF}

%In 5G NR, particularly at the mmWave bands, beamforming is considered as one of the key enablers for obtaining high performance communication links. Therefore, in the considered railway scenario, beamforming is utilized in both uplink and downlink directions. One considerable challenge in exploiting the full potential of beamforming %, is gruesome training procedures, which are 
%is the exhaustive training procedure required to find the optimum beam alignment between the transmitter and receiver. With multiple beam options at both sides of the communication link, the number of possible beam combinations increases exponentially. Fortunately, based on the reciprocity of the downlink and uplink radio channels, the same pair of RRH and train beams can be used in both directions, which is referred to as the beam correspondence in the 5G NR terminology. However,

In order to alleviate the complexity of the beam training procedure in the considered railway scenario, location-based beamforming is assumed at the RRH beamformers. Thus, each RRH adjusts the transmit and receive beams towards the estimated train position, which requires accurate and real-time train tracking capability in order to maintain the link connection in the considered high mobility scenario. Furthermore, it should be emphasized that the location-based beamformers in the RRHs are utilized for both the communications and positioning purposes. However, at the train side, where real-time position information is not directly accessible, the optimum %beamformer
beam selection is based on a conventional beam-sweeping-based training, where the train transmits a time-multiplexed set of SRSs over a predefined set of beams. %The received SRSs with different beams 
The SRSs with different transmit beams are then received by the RRHs, which are able to collect the positioning measurements and choose the best train-side beam for subsequent transmissions until the next set of SRSs are transmitted and the beams are updated.

The processing of the TDOA estimates from multiple RRHs builds on the extended Kalman filter (EKF), which includes joint tracking of the train position and velocity. For the sake of simplicity, the position and velocity in the considered scenario are defined in 2D coordinates, but the extension for supporting 3D coordinates, or including other tracked parameters, such as train acceleration, is straightforward. \red{Furthermore, although acceleration estimates would not be explicitly exploited for any purpose in the system, including acceleration in the EKF can sometimes improve the overall performance. However, for the studied HST scenario in this article, we did not observe any significant performance improvement by considering acceleration in the EKF, and thus excluded it from the model.} 

The fundamental operating principle of the EKF consists of two distinct steps, namely the prediction step and the measurement update step. %These steps are similar with the conventional Kalman filter, which is able to provide optimal tracking performance in scenarios with linear and Gaussian distributed prediction and measurement models. However, the utilized EKF enables an efficient solution in case of non-linear models via first-order Taylor series approximation. 
In the EKF processing, the state of the train, including the train position and velocity, is assumed to evolve based on a constant velocity model, where the train is assumed to travel with a constant velocity between two consecutive time steps. Moreover, the covariance of the corresponding state evolution process is based on a continuous white noise acceleration %(CWNA) 
model. 

\begin{figure}[t]
    \centering
    \includegraphics[width=0.95\columnwidth]{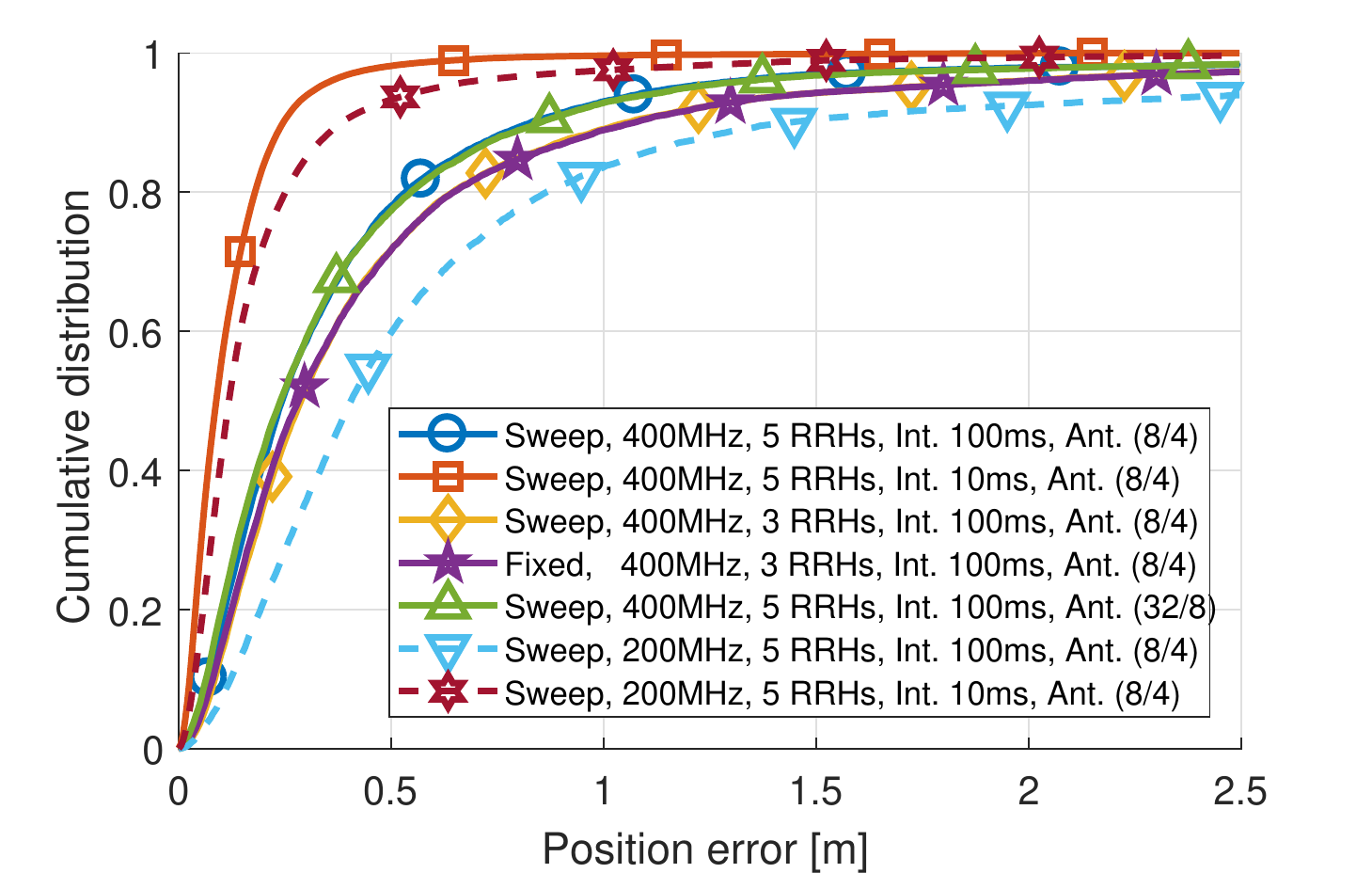}
    \caption{Cumulative distributions of the train position estimation error.}
    \label{fig:position_error}
\end{figure}

\begin{table}[t]
\renewcommand{\arraystretch}{1.2}
    \caption{Considered 5G NR physical layer parameterization} %\red{REDUCE accuracy as seen appropriate}} 
    \label{tab:PHYParams}
    \centering
    \begin{tabular}{|l|c|}
        \hline
        Parameter & Value \\
        \hline
        Carrier frequency & 30~GHz \\
        Channel bandwidth & [200,400]~MHz \\
        %Sampling rate & 491.52~MHz\\
        Subcarrier spacing & 120~kHz \\
        Allocation size  & [132,264]~PRBs \\
        Transmission rank & 2 (polarization based) \\
        %FFT size  & 4096 \\
        %CP length & 288 \\
        %Slot length & 14 OFDM symbols \\
        Channel model  & CDL-D 100~ns \\
        K-factor & 13.3~dB \\
        %Antenna elements per RRH panel & 8 \\ %(8,4,2,1,1)
        %Antenna elements per train panel & 4 \\ %(4,4,2,1,1)
        RRH antenna array size & 8 x 4 or 32 x 4 (hor. x ver.) \\
        Train antenna array size & 4 x 4 or 8 x 4 (hor. x ver.) \\
        Transmission power & +30~dBm \\
        %Evaluated distances to RRH & [10~m,290~m] \\
        Train velocity (max.) & 500~km/h \\
        Inter-RRH distance & 580~m \\
        Uplink SRS Tx interval & [10,100]~ms \\
        % &  18 and 24 (3GPP TS 38.214, MCS Table 2)\\
        %\multirow{2}{*}{Evaluated MCS indices} & 18 and 24 \\
        \multirow{2}{*}{Evaluated MCS indices} & 18 (64-QAM, R=822/1024) \\
       & 24 (256-QAM, R=841/1024)\\
        %Channel code & LDPC \\
        \hline
    \end{tabular}
    %\vspace{-3mm}
\end{table}

At predefined intervals, the train transmits uplink SRSs, which are time-multiplexed over the available beams. The SRSs are observed at multiple RRHs with different relative delays. Based on the cross-correlation between the received signal and the known SRS, timing measurements are obtained and compared with each other in order to assemble a set of TDOA measurements. %Hence, due to the relative nature of timing measurements, synchronization between the train and the network is not required. Moreover, considering that the RRHs are operating under the same BBU, synchronization between the RRHs is a reasonable and well-justified assumption.
The covariance of the TDOA measurements can, in turn, be estimated based on analytic Fisher information according to coarsely estimated signal-to-noise ratio (SNR) levels of each measurement. According to the TDOA principle, at least three RRHs are needed in order to obtain an unambiguous position estimate. %However, with the incorporated movement model in EKF, the position estimate can be updated also by using measurements from only two RRHs, but %obviously 
%with reduced positioning accuracy. 
% \begin{figure*}[!t]
%   \centering
%   \subfloat[][]{\includegraphics[angle=0,width=0.85\columnwidth]{epsFigs/commagTrain_10mDist_tput.eps}}
%   \qquad
%   \subfloat[][]{\includegraphics[angle=0,width=0.85\columnwidth]{epsFigs/commagTrain_290mDist_tput.eps}}
%   \caption{Link perf at 10 (BS/BS PN) and 290 m distances (BS/UE PN), NEED TO BE COMBINED INTO ONE FIGURE! Also 3dB SNR difference due to power scaling in the results! }
%   \label{fig:linkPerf}
%   %\vspace{-5mm}
% \end{figure*}

% \red{
% Possibly other stuff to mention based on the globecom paper \cite{GlobecomTrainPaper}, depends on how much space we have ):
% \begin{itemize}
%     \item Reference RRH selection
%     \item Outlier detection and removal due to NLOS detection and low-SNR measurements with fading
% \end{itemize}
% }
%The timing measurements at the RRHs are obtained by using location-based beamforming, where the receive beam of each RRH is directed towards the assumed train position. 
For obtaining each measurement, the beam direction is defined according to the predicted train position based on the previously updated position estimate and the included prediction model of the EKF. %Thus, by using the predicted train position, RRHs adjust their receive beams and obtain the timing measurements during each SRS transmission. After collecting the measurements, the train position estimate is updated and the succeeding SRS transmission is then received based on a new position prediction according to the current updated position. 
\red{Nevertheless, it should be noted that since obtaining high-quality positioning measurements and accomplishing accurately directed beams are interdependent, there is a possibility for error accumulation, especially if potential measurement outliers are not properly handled.}

\subsection{Positioning and Beam Alignment Performance}
\label{pos_performance}

The positioning performance is evaluated by using a simulated 100~km long train track identical to the one considered in \cite{GlobecomTrainPaper}, but with slight additional curvatures along the route. In the beginning, the train is standing still and begins to accelerate with full power. After reaching the maximum velocity of 500~km/h the train maintains its velocity for about 4 minutes. After this, the train slows down to around 290 km/h velocity, but re-accelerates again, until stopping in the end. The number of RRHs taking simultaneous timing measurements is considered to be either three or five. Detailed parameterization of the underlying 5G NR physical layer is shown in Table \ref{tab:PHYParams}, where the signal bandwidth and the number of antenna elements are varied between different positioning performance results. \red{Similar to the modeling in \cite{GlobecomTrainPaper}, the radio propagation related path loss model and the corresponding spatially correlated shadowing model are determined according to the 3GPP-specified urban micro model characteristics. As shown in Table \ref{tab:PHYParams}, the considered fast fading channel model in this article is the 3GPP-specified time varying Clustered Delay Line D (CDL-D) with a dominating LOS path. Moreover, since the utilized TDOA-based positioning approach presumes ranging via LOS paths, the obtained performance results are highly dependent on the LOS path availability. However, when the LOS path availability is compromised, for example, when another train is blocking the signal, appropriate LOS detection methods and measurement outlier detection methods, such as the ones presented in \cite{GlobecomTrainPaper}, can be used to mitigate positioning errors due to reflected paths. Nevertheless, in the considered scenario, where multiple RRHs are assumed to be located withing the TDOA measurement range, the performance degradation due to occasional LOS signal blocking is expected to be minor.}

In Fig. \ref{fig:position_error}, the cumulative distribution of the train position estimation error is shown with variable measurement and EKF update intervals, antenna configurations, bandwidths, and uplink beamforming strategies. Regarding the latter one, besides the earlier described beam sweeping based SRS transmissions at the train side, we consider as an alternative using also fixed beams at the train as a reference \cite{GlobecomTrainPaper}. In the case of fixed beams, the train transmits only a single SRS with the beams pointing towards the nose and the tail of the train. Depending on the beamwidth, the link quality in proximity of a RRH might suffer due to the poor beam alignment. Furthermore, the antenna panels at the RRHs and the train comprise a uniform rectangular array, where the number of vertical direction antenna elements is fixed to 4. \red{However, the number of horizontal direction antenna elements is varied in the evaluations, and it is either 4 or 8 at the train side, and either 8 or 32 at the RRH side. Moreover, it should be noticed that the required angular resolution of the uplink beam sweeping procedure depends on the considered beamwidth, and thus, the number of antenna elements at the train side.}

\begin{figure}[t]
    \centering
    \includegraphics[width=0.95\columnwidth]{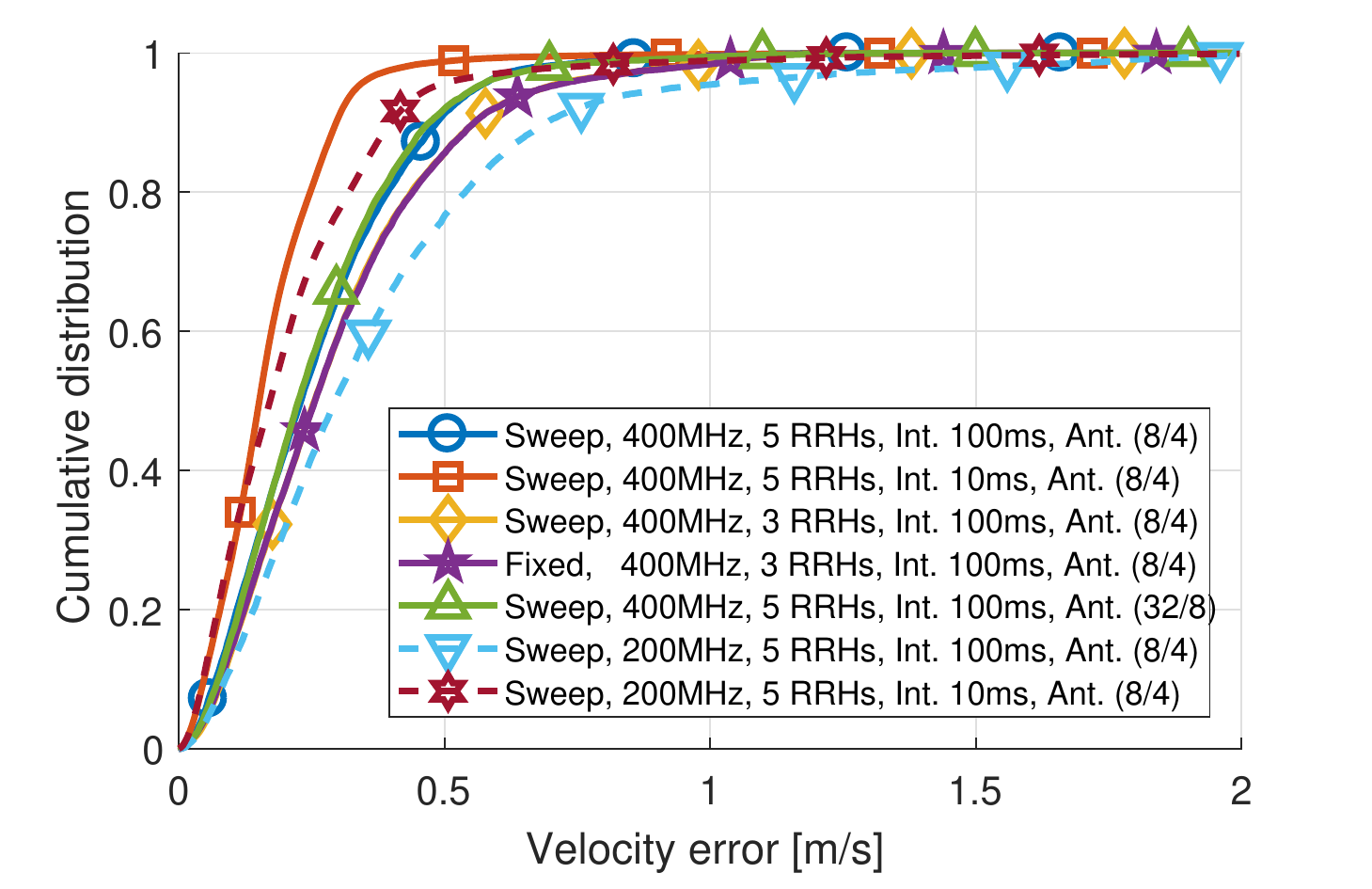}
    \caption{Cumulative distributions of the train velocity estimation error.}
    \label{fig:velocity_error}
\end{figure}

\begin{figure}[t]
    \centering
    \includegraphics[width=0.95\columnwidth]{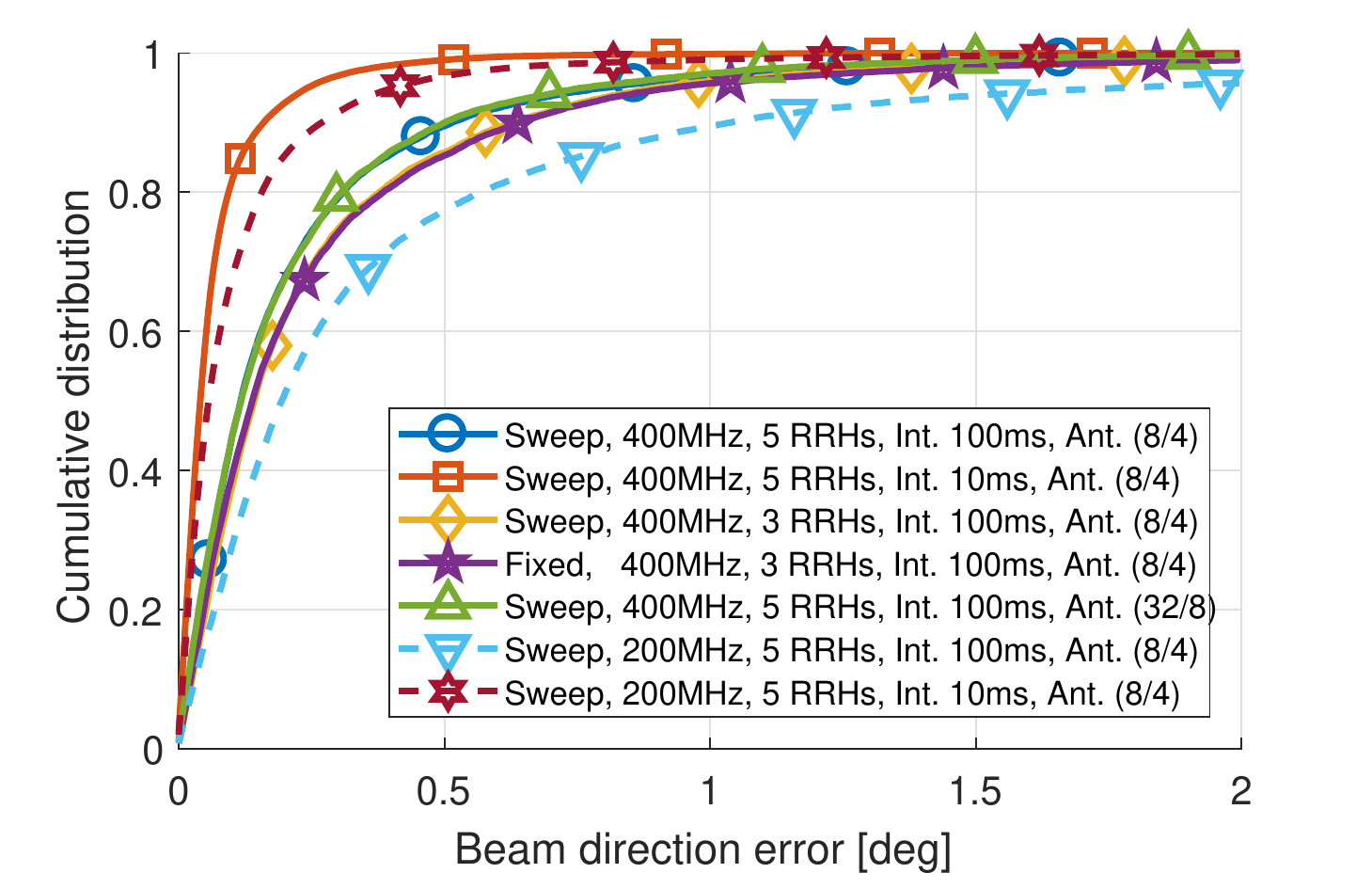}
    \caption{Cumulative distributions of the RRH beam direction error with the location-based beamforming, when obtaining the position measurements.}%The results include the beam direction error only for the closest RRH at a time.}
    \label{fig:beam_error}
\end{figure}

As seen in Fig. \ref{fig:position_error}, the positioning accuracy is considerably improved when using 10~ms ("Int. 10~ms") measurement update interval compared to the 100~ms ("Int. 100~ms") interval. However, neither the number of antenna elements ("Ant., number of horizontal elements at RRH / at train") nor the beamforming strategy (i.e. "Fixed" or "Sweep" beamformer at the train side) have a considerable effect on the positioning accuracy. \red{On the contrary, the performance with the 400~MHz bandwidth (i.e. 264 physical resource blocks, PRBs) is considerably better compared to the 200~MHz bandwidth (i.e. 132 PRBs), stemming from the increased TDOA ranging resolution.} In addition, as expected, the position estimation accuracy is improved, when using measurements from 5 RRHs instead of only 3 RRHs. Similar observations can be done, when considering the train velocity estimation error, whose cumulative distributions are shown in Fig. \ref{fig:velocity_error}. %Again there is a clear difference when comparing the results over different bandwidths. However, the performance improvement with the 10~ms measurement update interval compared to the 100~ms interval has been slightly decreased.

With location-based beamforming, the beam direction error is a crucial performance indicator, as it determines the achievable beam gains, but also the possible beam losses in case of substantial beam misalignment. In Fig. \ref{fig:beam_error}, the cumulative distributions of the RRH beam direction error are shown at the time instants where the positioning measurements are taken. In these error curves, only the beam error of the closest RRH is considered at the time, since due to the specific system geometry, RRHs further away have generally smaller beam direction errors. As the beam direction accuracy depends on the positioning accuracy, the results are consistent with the results shown in Fig. \ref{fig:position_error}.

%Due to the specific system geometry, RRHs further away from the train have generally smaller beam direction errors, and thus, considering those would only improve the overall beam direction accuracy. Moreover, since the beam direction accuracy depends on the accuracy of the train position estimate, the results are consistent with the results shown in Fig. \ref{fig:position_error}. %Again, using a higher bandwidth, a higher measurement update interval, or a higher number of RRHs for taking the measurements, improves the estimation performance, whereas the effect of other parameters is practically negligible. %\red{ToLe: Here an example of the beamforming gain loss with 32 antenna elements in the RRH with 99\% reliability alignment error could be given, if space} 

\begin{figure*}[!t]
  \centering
  \includegraphics[width=0.9\textwidth]{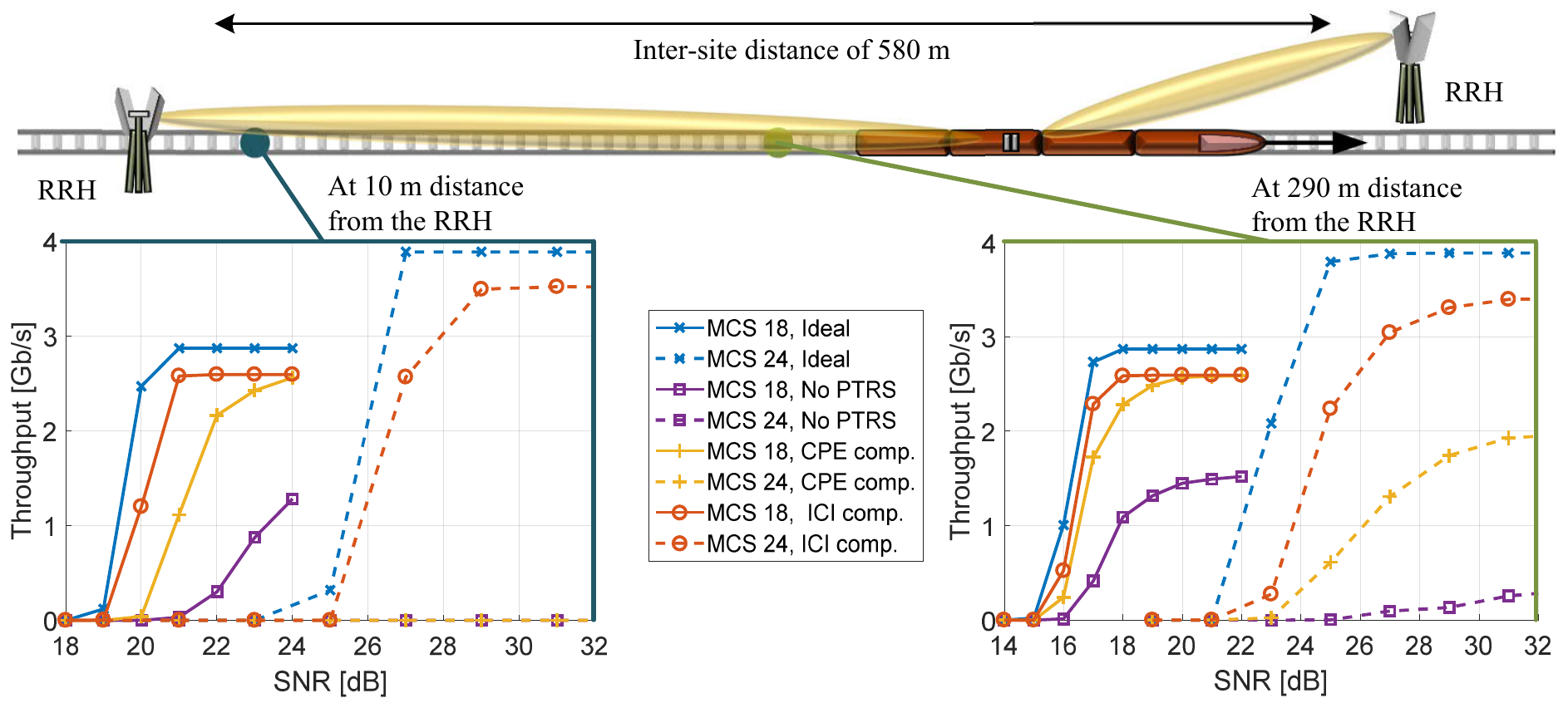}
  \caption{Throughput of the HST communications link as a function of the SNR with different parameterizations at example distances of 10~m and 290~m from a RRH.}
  \label{fig:linkPerf}
  %\vspace{-5mm}
\end{figure*}

\section{5G NR Communications for Modern Railways}
\label{sec:5GComms}

% \red{
% DESCRIPTION OF THE CONSIDERED COMMUNICATIONS APPROACH AND RELATED RESULTS

% \begin{itemize}
%     \item Overall description of the considered communications approach
%     \item Doppler compensation principle
%     \item Used beamforming strategy
%     \item Comms part results (Doppler compensation)
% \end{itemize}
% }

In this section, the throughput performance of the 5G NR based HST radio link, building on the technical enablers described in Section \ref{sec:5GProspects}, is demonstrated. The main evaluation parameters are similar to the positioning evaluations, shown in Table \ref{tab:PHYParams}, and follow the general high speed train scenario evaluation assumptions defined in \cite{3GPPTR38802}. \red{The 400~MHz channel bandwidth case is assumed, while the antenna array sizes are $8\text{x}4$ (RRHs) and $4\text{x}4$ (train).} Specifically, we assume that the network uses the position and velocity estimates to calculate the Doppler frequency and timing alignment precompensation factors as described in Section \ref{sec:5GProspectsComms}. 

In addition to the advanced Tx processing in the form of precompensation, a novel block PTRS is used to allow PN or residual Doppler induced ICI compensation. As described earlier, the NR Rel-15 PTRS is designed to only allow for the estimation and compensation of the CPE introduced by PN and modest frequency offset. %The actual structure of the PTRS relies on transmitting known symbols on individual subcarriers repeated in every two or four PRBs in frequency domain and in every, every second, or every fourth OFDM symbol in time domain. As the Rel-15 PTRS is distributed in the frequency domain, only CPE estimation is possible.
%
%On the other hand, the novel block PTRS design relies on contiguous allocations in frequency domain and the size of the block PTRS is assumed to be defined in multiples of PRBs% or 12 subcarriers. 
%
The more elaborate contiguous block PTRS structure allows for the estimation of the ICI components by solving a set of linear equations limited by the number of reference symbols in the block PTRS as described in \cite{VillesPhaseNoisePaper}. In the presented results, we have assumed a block PTRS of size %48 subcarriers
4~PRBs to be transmitted in %each spatial layer
both spatial layers and we have configured the 5G NR Rel-15 compliant PTRS sturcture to use similar amount of subcarriers to provide equal overhead for the two reference signals structures. Additionally, a 3GPP compliant PN model is assumed. 

As shown in Fig. \ref{fig:linkPerf}, the downlink throughput performance of the radio link is evaluated at two different distances from the closest RRH, namely 10~m and 290~m distance, \red{for varying SNR levels such that different MCSs can be show-cased}. %These example points demonstrate the operation of the radio link near by a RRH and between two RRHs. 
Regarding the communications link, it is here assumed that the train is always connected to two RRHs in a single frequency network. The "Ideal" reference curves demonstrate the link performance without PN, position or velocity estimation errors, or beam alignment errors. All other cases include these non-idealities, and it is assumed that the errors follow the statistics presented in Section \ref{pos_performance}. %It should be noted that especially the Doppler precompensation effect is significant - that is, without Doppler precompensation the communications link does not work with the evaluated MCS indices yielding zero throughput. 
The "No PTRS" case corresponds to performance without PTRS and the "CPE comp." case corresponds to 5G NR Rel-15 compliant distributed PTRS structure which allows to estimate and compensate only the CPE in the received
signal. Finally, the "ICI comp." case refers to the scenario where the block PTRS structure is enabled allowing the receiver to estimate and compensate also for the ICI induced by the PN and residual Doppler error effects. 

% \begin{figure}[t]
%     \centering
%     \includegraphics[width=\columnwidth]{epsFigs/commagTrain_10mDist_tput.eps}
%     \caption{Throughput}
%     %\label{fig:pos_error_all}
% \end{figure}

We can observe that with 64-QAM modulation (MCS~18) the distributed PTRS provides clear performance improvement in the throughput performance compared to the "No PTRS" case, while block PTRS is able to achieve the maximal throughput at lower SNR. In the case of 256-QAM modulated data signal (MCS~24), the block PTRS shows significant gain over the distributed PTRS allowing to achieve 3.5~Gbps throughput \red{at both the 10~m and 290~m distances, whereas with the distributed PTRS the radio link does not work at the 10~m distance.} This highlights the benefits of block PTRS, not only for PN distortion compensation, but also for residual Doppler error compensation. Without any kind of PTRS, the radio link does not work with 256-QAM modulation. It can be concluded that the considered transmitter pre-compensation and receiver processing techniques significantly improve the link performance and allow to obtain ultra high throughput in the HST scenario.

\section{Conclusion}
\label{sec:conclusions}

The emerging 5G networks with flexible radio interface design open new opportunities for modern railway systems, including both extreme broadband communication links and various mission critical railway management services. It was shown that with 3GPP-specified 5G NR parametrization, when using the 400~MHz bandwidth and TDOA measurements processed through an EKF with 10ms interval, it is possible to achieve submeter positioning accuracy % with as high as
over 99\% of the time. Moreover, by exploiting the obtained position information in the underlying 5G NR communications system for precompensating the Doppler shift and timing alignment, combined with novel block PTRS based receiver algorithms, up to 3.5~Gbps throughput %was reached with the highest considered MCS index. 
can be supported throughout the track. Besides the Doppler shift precompensation and timing alignment, position information was used for location-based beamforming at the RRHs to simplify the beam training processes. Hence, with the proposed high-efficiency positioning approach and the related utilization of location-awareness in the physical layer processing methods and location-aware RRM, 5G networks have the potential to revolutionize the communications and management systems for modern railways.

% \section*{Acknowledgment}
% \red{use footnote...}

% \ifCLASSOPTIONcaptionsoff
%   \newpage
% \fi

\bibliographystyle{IEEEtran}
\bibliography{IEEEabrv,main}

\section*{Biographies}

\vspace*{-13\baselineskip}
\begin{IEEEbiographynophoto}{Jukka Talvitie} is a university lecturer at Tampere University, Finland. His research interests include signal processing for wireless communications, and network-based positioning methods with particular focus on 5G networks.
%received the M.Sc. and D.Sc. degrees from Tampere University of Technology, Finland, in 2008 and 2016, respectively. He is currently a University Lecturer with the \red{Department} of Electrical Engineering at Tampere University. His research interests include signal processing for wireless communications, and positioning in 5G networks.
\end{IEEEbiographynophoto}

\vspace*{-13\baselineskip}
\begin{IEEEbiographynophoto}{Toni Levanen} is a post-doctoral researcher at Tampere University, Finland. His current research interests include physical layer design for 5G NR, interference modeling in 5G cells, and high-mobility support in mmWave communications.
%received the M.Sc. and D.Sc. degrees from Tampere University of Technology, Finland, in 2007 and 2014, respectively. He is currently with the the \red{Department} of Electrical Engineering at Tampere University. %In addition to his contributions in academic research, he has worked in industry on wide variety of development and research projects. 
%His current research interests include physical layer design for 5G NR, interference modeling in 5G cells, and high-mobility support in mmWave communications.
\end{IEEEbiographynophoto}

\vspace*{-13\baselineskip}
\begin{IEEEbiographynophoto}{Mike Koivisto} is a researcher at Tampere University, Finland. His research interests include positioning, with an emphasis on network-based positioning and the utilization of location information in mobile networks.
%[S’16] (mike.koivisto@tut.fi) received the M.Sc. degree in mathematics from Tampere University of Technology, Finland, in 2015, where he is currently pursuing the Ph.D. degree. %From 2013 to 2016, he was a research assistant with TUT. 
%He is currently a researcher with the \red{Department} of Electrical Engineering at Tampere University. His research interests include positioning, with an emphasis on network-based positioning and the utilization of location information in mobile networks.
\end{IEEEbiographynophoto}

\vspace*{-13\baselineskip}
\begin{IEEEbiographynophoto}{Tero Ihalainen} is a senior specialist at Nokia Bell Labs, Finland. His research interests include signal processing for radio communications, physical layer design of 5G New Radio, and techniques and algorithms supporting high-mobility in millimeter-wave communications.
%received the M.Sc. degree in electrical engineering and the Dr.Tech. degree in communications engineering from Tampere University of Technology, Finland, in 2005 and 2011, respectively. He is currently with the Nokia Bell Labs. His research interests include signal processing for radio communications, physical layer design of 5G New Radio, and techniques and algorithms supporting high-mobility in millimeter-wave communications.
\end{IEEEbiographynophoto}
% received the M.Sc. degree in electrical engineering and the Dr.Tech. degree in communications engineering from Tampere University of Technology (TUT), Finland, in 2005 and 2011, respectively. From 2005 to 2010, he was a Researcher with the Department of Communications Engineering, TUT. From 2011 to 2014, he was with Nokia Research Center, Finland. Currently, he is with the Nokia Bell Labs. His research interests include signal processing for radio communications, physical layer design of 5G New Radio, and techniques and algorithms supporting high-mobility in millimeter-wave communications.

\vspace*{-13\baselineskip}
\begin{IEEEbiographynophoto}{Kari Pajukoski} is a Fellow ate Nokia Bell Labs, Finland. His research interests include cellular standardization, link and system simulation, and algorithm development.
%received his B.S.E.E. degree from the Oulu University of Applied Sciences in 1992. He is a Fellow with the Nokia Bell Labs. He has a broad experience from cellular standardization, link and system simulation, and algorithm development. He has more than 100 issued US patents, from which more than 50 have been declared “standards essential patents”, and is author or co-author of more than 300 standards contributions and 30 publications.
\end{IEEEbiographynophoto}

\vspace*{-13\baselineskip}
\begin{IEEEbiographynophoto}{Mikko Valkama} is a Full Professor at Tampere University, Finland. His research interests include radio communications, radio systems and signal processing, with specific emphasis on 5G and beyond mobile networks.
%[S'00, M'02, SM'15] (mikko.e.valkama@tut.fi) received his M.Sc. and D.Sc. degrees (both with honors) from Tampere University of Technology, Finland, in 2000 and 2001, respectively. %In 2003, he worked as a visiting researcher at San Diego State University, California. 
%Currently, he is a Full Professor and Head of the \red{Department} of Electrical Engineering at Tampere University. His research interests include radio communications, radio systems and signal processing, with specific emphasis on 5G and beyond  mobile networks.
\end{IEEEbiographynophoto}

\end{document}